\documentclass{aa}  

\usepackage{graphicx}
\usepackage{lineno}
\usepackage{txfonts}
\usepackage[normalem]{ulem}

\begin{document} 

   \title{Closing the loop on-sky with a vector-Zernike wavefront sensor using a convolutional neural network as phase reconstructor}
   \titlerunning{Closing the loop on-sky with a ZWFS using a CNN as phase reconstructor}

   \author{F. Oyarzún
          \inst{1},
          M. Motte  \inst{2,1},
          C. Heritier \inst{2,1},
          B. González \inst{4},
          R. Muñoz \inst{4},
          V. Chambouleyron \inst{1},
          M.A. Alagao \inst{1},
          A. Striffling \inst{1},
          M. Pasinetti \inst{1},
          E. Vinerskas \inst{1},
          P. Trouve-Peloux \inst{3},
          F. Champagnat \inst{3},
          T. Fusco \inst{2,1},
          \and
          B. Neichel \inst{1}
          }

   \institute{Aix Marseille Univ, CNRS, CNES, LAM, Marseille, France\\
              \email{francisco.oyarzun@lam.fr}
         \and
             DOTA, ONERA, Université Paris Saclay, F-91123 Palaiseau, France
        \and
            DTIS, ONERA, Université Paris Saclay, 91123 Palaiseau, France
        \and
            School of Electrical Engineering, Pontificia Universidad Católica de Valparaíso, Valparaíso, Chile
            }

   \authorrunning{F. Oyarzún}
   \date{\today}

  \abstract
   {The new giant segmented mirror telescopes will use adaptive optics to reach the fundamental limits in resolving power. To accomplish this, new wavefront sensors (WFS) have been designed to fulfill the requirements, but they may require non-linear reconstruction techniques to operate given the nature of the signal of the WFSs}
   {In this article we show that it is possible to use non-linear reconstructors to extend the dynamic range of one of the most sensitive wavefront sensors far beyond the designed limits: the Zernike wavefront sensor (ZWFS).}
   {We trained a convolutional neural network (CNN) completely in simulation to perform the phase reconstruction of a vector-ZWFS (v-ZWFS), a higher dynamic-range variant of the ZWFS. Contrary to the linear method, the CNN uses the information across the full frame to reconstruct the phase at each point, enabling it to resolve the ambiguities introduced by the periodic response of the ZWFS and thereby extend its effective capture range. We developed a two-step training strategy that ensured closed-loop stability and used a physically informed loss function to maximize the performance of the CNN.}
   {We successfully closed the loop on-sky with the v-ZWFS using the CNN in observing conditions that the linear reconstructor could not converge to a stable flat wavefront. In cases where both reconstruction methods were working, the CNN outperformed the linear method in almost all cases, and in favorable seeing conditions we were even able to close the loop with the ZWFS acting as a first stage WFS, highlighting the extended dynamic range brought by the use of a non-linear wavefront reconstructor.}
   {We conclude that the use of non-linear wavefront reconstructor can extend the use cases of adaptive optics systems, especially when the WFS used shows highly non-linear behaviors.}

    
   \maketitle
   \nolinenumbers
%
%-------------------------------------------------------------------

\section{Introduction}

Direct imaging and atmospheric characterization of exoplanets will be one of the main objectives of the new giant telescopes \citep{2021Msngr.182...38K}. Nevertheless, ground-based telescopes suffer from resolution loss due to atmospheric turbulence. Instead of achieving the diffraction-limited resolution \( \theta_{\text{diffraction}} = \lambda / D \), the effective resolution is degraded to \( \theta_{\text{seeing}} = \lambda / r_0 \), where \( r_0 \) is the Fried parameter \citep{1966JOSA...56.1372F}, typically ranging from 3 cm to 15 cm at $500 \, nm$ in good observatories. This results in a resolution loss of 50 to 250 times for an 8 m telescope, and up to 1000 times for a 40 m one. Adaptive optics (AO) systems address this limitation in real time, using a wavefront sensor (WFS) to measure aberrations, a deformable mirror (DM) to correct them, and a control system to compute the corrections.

A WFS transforms phase distortions \(\phi(x,y)\) into an intensity signal $I(\phi)$ measurable by a detector. Fourier Filtering WFSs (FFWFS), such as the Zernike (ZWFS, \cite{2013A&A...555A..94N}) or Pyramid WFSs (PWFS, \cite{1996JMOp...43..289R}) are particularly attractive for their high sensitivity \citep{2004OptCo.233...27V, 2023A&A...670A.153C}, enabling observations of fainter guide stars and increasing sky coverage \citep{2018ARA&A..56..315G}. FFWFSs relate the input phase and the intensity in the detector with the following non-linear relation \citep{2016Optic...3.1440F}:

\begin{equation}
    I(\phi) = \left|\mathcal{F}\{\mathcal{F}\{\mathbb I_p e^{i\phi}\} \cdot m \} \right|^2 = \left|\left( \mathbb I_p e^{i\phi} \right) \ast \mathcal{F}\{m\} \right|^2,
    \label{eq:FFWFS_general_equation}
\end{equation}

where \(\mathcal{F}\{\}\) is the Fourier transform, $\mathbb I_p$ the pupil indicative function, $\ast$ the convolution operation and $m$ is a complex-valued filtering mask. While this relation can be linearized for small phase amplitudes, the nonlinear response at larger aberrations limits the performance of standard reconstruction techniques, and may require advanced reconstruction techniques such as iterative solvers \citep{2024A&A...681A..48C} or machine learning methods \citep{2020OExpr..2816644L,2024A&A...687A.202W}.

Neural networks (NNs) have shown promise as wavefront reconstructors. Previous studies have used NN to reconstruct the signal from a PWFS \citep{2020OExpr..2816644L,2023PASP..135k4501W, 2024A&A...687A.202W,Tang:26}, with the first on-sky demonstration by \citet{2025A&A...696L...1L}. To first order, the non-linearities of the PWFS manifest as a gain saturation, meaning that while the amplitude of the reconstruction may be underestimated, the sign is always correct. As a result, in closed loop each iteration still corrects in the right direction, and the nonlinearity effectively acts as a dynamic gain that can be compensated for \citep{2021A&A...649A..70C, 2025A&A...703A.253S}, meaning that traditional linear reconstruction techniques could work in most scenarios.

In this work, we focus on the Zernike wavefront sensor, a Fourier filtering sensor that offers higher sensitivity than the PWFS \citep{2021A&A...650L...8C}. The ZWFS consists of a phase-shifting dot placed at the center of the focal plane, which introduces a phase shift $\delta$ (usually $\pi/2$) to the core of the point spread function (PSF). The size of the dot $\rho$ is typically chosen to be between 1 and 2 $\lambda/D$, where $\lambda$ is the sensing wavelength and $D$ the telescope diameter.

However, the main challenge of the ZWFS is that its output signal contains a constant offset plus terms proportional to the sine and cosine of the input phase $I(\phi) \sim \alpha \sin(\phi) + \beta\cos(\phi) + \gamma$, meaning that for large phase amplitudes the periodic response of the ZWFS causes the measured signal to wrap, introducing ambiguities in the phase estimate. Unlike the PWFS nonlinearity discussed above, this wrapping can produce estimates of the wrong sign, potentially driving the control loop in the wrong direction and causing instabilities. Traditional linear reconstructors are therefore limited to regimes where the residual phase is small enough to avoid wrapping.

The achievable capture range depends on both the depth and diameter of the Zernike dot: the depth sets the relative sine/cosine weighting of the signal, and hence the bijective range for the inversion. The dot diameter, in turn, sets the spatial cut-off frequency of the reference wave used in the linear model; for larger dots, this reference is no longer independent of the input phase, introducing a bias that the standard linear reconstructor does not account for \citep{2020Optic...7.1267S}.

To mitigate this issue, a variant of the ZWFS was introduced, called vector-ZWFS (v-ZWFS), originally proposed by \citet{2019OptL...44...17D} to measure amplitude variations, and later by \citet{2022SPIE12185E..0XC} and \citet{2024ApJ...967..171S} as a larger dynamic-range version of the ZWFS. The v-ZWFS is composed of two Zernike masks with different phase shifts, such that the signal from each has different proportions of sine and cosine, which can double the linear range of the sensor, at the cost of a reduced sensitivity given the larger number of pixels needed to capture the two pupils. By computing linear combinations of the two signals one can isolate the pure sine and cosine components and then the phase reconstruction can be done by taking the $\arctan$ of the result, effectively doubling the dynamic range \citep{2022SPIE12185E..0XC}. However, this approach still fails when aberrations are large enough that even the extended range is exceeded.

The limitations of the v-ZWFS motivate the use of a more powerful reconstruction approach. In this work, we propose using a Convolutional Neural Network (CNN) to reconstruct the phase from the v-ZWFS signals, simultaneously performing phase reconstruction and unwrapping in a single step. This is possible because, unlike linear reconstructors that rely on local information, a CNN can leverage the full spatial context of the detector image to reconstruct the phase at each point, making it naturally suited to handle the wrapping nonlinearity. To our knowledge, this is the first application of a CNN reconstructor to the ZWFS, extending the deep learning approach that has shown success on the PWFS to a sensor with a fundamentally more challenging nonlinearity. 

We demonstrate in this study a on-sky closed loop using CNN on OZIRIIS \citep{oziriis}, the second-stage AO system of the PAPYRUS \citep{papyrus} instrument at OHP, which is equipped with a v-ZWFS in the H-band (1.5 - 1.7 $\mu m$) and a 97-actuators DM. We train the CNN reconstructor entirely in simulation, modeling the key components of the AO system and generating a dataset broad enough to ensure generalization across a wide range of observing conditions, and demonstrate its performance on-sky. The simulation setup, training strategy, and on-sky demonstration are described in Sections \ref{sec:methods} and \ref{sec:results}.

\section{Methods}
\label{sec:methods}

\subsection{Modeling of the wavefront sensor}
An accurate forward model of the WFS is required both to generate training data and to compute the synthetic reference frame used in preprocessing. Following Eq. \ref{eq:FFWFS_general_equation}, the signal of the ZWFS can be obtained via two successive Fourier transforms and by setting the filtering mask to be

\begin{equation}
    \begin{split}
          m(x,y) = e^{i\delta\Delta(x,y)} \\
        \Delta(x,y) = \begin{cases}
                1 & \text{if } x^2 + y^2 \leq \rho^2\\
                0 & \text{if } x^2 + y^2 > \rho^2.
        \end{cases}
    \end{split}
\end{equation}

\begin{figure}
    \centering
    \includegraphics[width=0.99\linewidth]{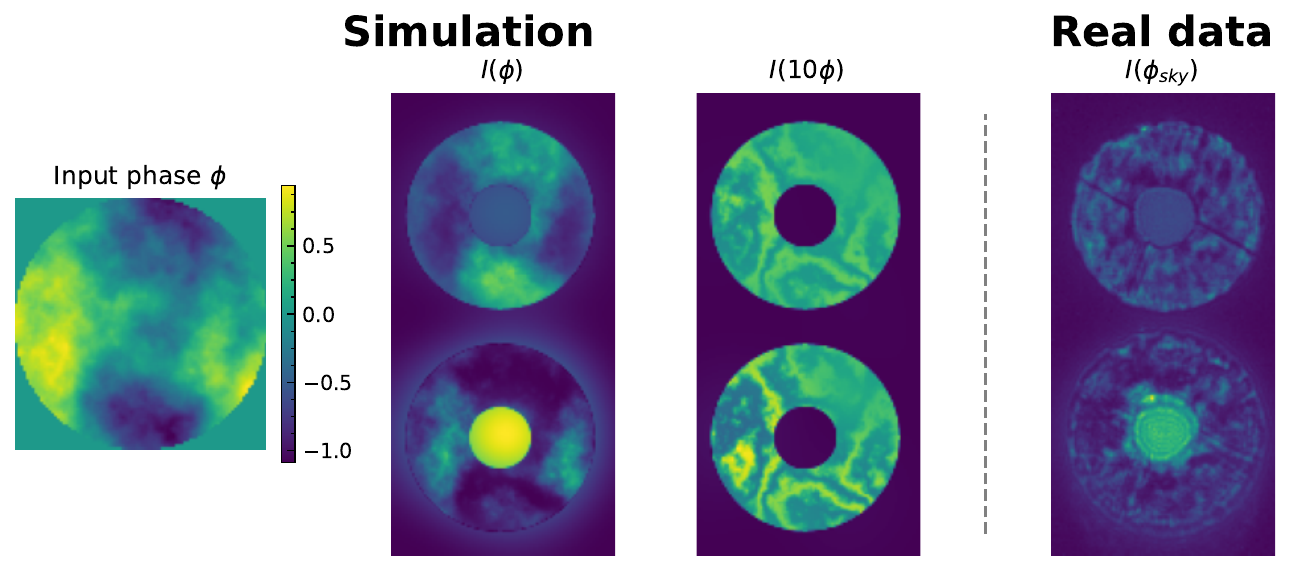}
    \caption{Example of the signal obtained with a v-ZWFS on simulation and on-sky. The top pupil corresponds to the Zernike mask with a phase shift of $\delta_2 = -0.8\pi$ and the bottom pupil to the mask with a phase shift of $\delta_1 = 0.3\pi$. Simulation: the left panel shows the input phase $\phi$, the middle panel is the signal from the v-ZWFS for an input $\phi$, and the right panel shows the signal for an input ten times larger. Real data: the panel shows an on-sky frame from the OZIRIIS instrument.}
    \label{fig:vZWFS_signal}
\end{figure}

The signal of the v-ZWFS was then obtained by two separate and incoherent propagations and setting the appropriate phase shift $\delta$ for each mask. For OZIRIIS, both masks share the same dot size of 2.15 $\lambda/D$, with phase shifts $\delta_1 = 0.3\pi$ and $\delta_2 = -0.8\pi$ radians. The physical implementation in the instrument is achieved using a meta-surface mask as in \citet{Wenger:25}. Figure \ref{fig:vZWFS_signal} shows the signal obtained when using a v-ZWFS both in simulation and from the instrument on-sky. 

For small values of phase ($\phi$, middle image), the v-ZWFS has a linear response to phase, whereas for large values ($10\phi$, right image) the signal wraps, which can be observed as the contour-like lines that show in the bottom part of the bottom pupil, where different input phases produce similar measured intensities. If using a linear reconstructor, the AO loop would diverge for the second scenario. The on-sky pupils (even though the phase is different) share similarities with the later case on the simulation, highlighting the accurate modeling of the WFS. Diffraction features from the telescope spiders, visible in the on-sky bench images, were not included in the WFS simulation. This simplification did not noticeably affect performance, as no degradation was observed in the reconstruction near the spider regions.

\subsection{Modeling of the deformable mirror}

To simulate OZIRIIS it was necessary to introduce the misregistration of the DM with respect to the WFS. This required accounting for the possible translational offsets, rotations and magnification differences between the DM actuators and the WFS subapertures. Unaccounted misregistration leads to model mismatch, degraded reconstruction performance and possible instabilities during closed loop, all of which negatively impact the operation of the instrument \citep{2018MNRAS.481.2829H}. 
To compute the misregistrations we used the SPRINT algorithm developed by \citet{2021MNRAS.504.4274H}, in which we needed an accurate model of the WFS and one experimentally measured interaction matrix using OZIRIIS. By comparing the experimental to synthetic interaction matrices, the SPRINT algorithm iteratively fits the required degrees of freedom of the DM to account for the misregistrations. Once the DM model was established, the first 50 (out of 87) Karhunen-Loève (KL) modes were computed as the modal basis for training. This number was chosen as a practical trade-off: in simulation, including higher-order modes yielded marginal improvement in correction performance while significantly increasing the risk of closed-loop instability.

\subsection{Frame preprocessing} 

In the OZIRIIS instrument each pupil was 90 pixels across (9 times oversampling the DM) on a 256 $\times$ 128 frame, so as a preprocessing step we (1) extracted the pupil images from the detector frame; (2) subtracted a previously computed synthetic reference frame that serves as both the zero-phase reference \citep{2016Optic...3.1440F} and already bringing the input to the network to zero mean; and (3) normalized the images by a fixed scalar such that the standard deviation of the pupil images is unity on average across the dataset. This normalization stabilizes training by keeping input amplitudes within a consistent range regardless of flux level.

The reference frame was computed synthetically from the WFS model rather than measured on-sky, since no direct access to a diffraction-limited on-sky reference is available with OZIRIIS. To do so, we first sent a strong tilt to the DM to separate the PSF from the mask and recover the pupil indicative function $\mathbb I_p$. This pupil was then propagated through the WFS model (Eq. \ref{eq:FFWFS_general_equation}) assuming a flat wavefront ($\phi = 0$), and the resulting intensity was used as the synthetic reference frame.

\subsection{Neural network reconstructor}

We used a CNN for wavefront reconstruction. Given that the v-ZWFS produces two pupil images, the input to the network consisted of two channels, one for each pupil. The preprocessed images were passed to a CNN with the architecture detailed in Tab. \ref{tab:cnn_architecture}: the network consisted of six convolutional layers with 16 to 512 filters with kernel sizes from 11$\times$11 to 3$\times$3, applying leaky ReLU activation and $2\times2$ max-pooling for downsampling after each layer. A final adaptive average pooling layer extracts global features, followed by a fully connected layer that maps them to the first \(N = 50\) modal coefficients. The selected architecture was optimized on simulation, balancing the size of the network for estimation accuracy and inference time. The final network architecture had $\sim$ 1 million parameters, and after compilation to a TensorRT engine the inference could be performed in 260 $\mu s$ using an NVIDIA A40 GPU. An extra 100 $\mu s$ of preprocessing time led to a total of 360 $\mu s$ for phase reconstruction.

\begin{table}
    \centering
    \caption{CNN architecture.}
    \label{tab:cnn_architecture}
    \begin{tabular}{lcccc}
    \hline\hline
    Layer & In $\rightarrow$ Out ch. & Kernel & Padding & Size after pool \\
    \hline
    Conv1 & 2 $\rightarrow$ 16   & 11$\times$11 & 7 & $H/2$ \\
    Conv2 & 16 $\rightarrow$ 32  & 7$\times$7   & 5 & $H/4$ \\
    Conv3 & 32 $\rightarrow$ 64  & 5$\times$5   & 3 & $H/8$ \\
    Conv4 & 64 $\rightarrow$ 128 & 3$\times$3   & 2 & $H/16$ \\
    Conv5 & 128 $\rightarrow$ 256& 3$\times$3   & 2 & $H/32$ \\
    Conv6 & 256 $\rightarrow$ 512& 2$\times$2   & 2 & $H/64$ \\
    \hline
    \multicolumn{5}{l}{Adaptive average pooling $\rightarrow$ (1,1), Dropout ($p=0.1$)} \\
    \multicolumn{5}{l}{Fully connected: $512 \rightarrow N_\mathrm{modes}$} \\
    \hline
    \end{tabular}
    \tablefoot{
    All convolutional layers use LeakyReLU activation and are followed by 2$\times$2 max-pooling (stride 2). Input: two-channel pupil image.
    }
\end{table}

\subsection{Dataset}

To generate the dataset to train the system, we exploited the known power spectral density (PSD) of atmospheric wavefronts, allowing us to generate an arbitrarily large number of samples \citep{2023aoel.confE..50T}. The PSD of the wavefront can be constructed by combining three main sources: the PSD of the atmosphere, which follows a Kolmogorov or Von-Karman distribution \citep{1991RSPSA.434....9K}, the PSD of the deformable mirror, modeled as unity outside the correction radius and, inside it, scaled by a factor drawn uniformly from [0,1] for each sample to represent the continuum between open-loop, intermediate, and closed-loop residual statistics \citep{1998SPIE.3353.1038R}, and the PSD of the temporal errors arising from the finite AO loop latency and wind-driven evolution of the atmosphere \citep{1998SPIE.3353.1038R}. To generate each wavefront sample, we drew a grid of complex Gaussian random variables in the spatial frequency domain. This random field is then filtered by the square root of the combined PSDs, which include contributions from the atmospheric turbulence, deformable mirror dynamics and temporal lag. The wavefront sample can be obtained as the real (or imaginary) part of the inverse Fourier transform of the resulting spectrum. Once a wavefront is generated, it is propagated through the WFS to produce the input image for the neural network. The dataset contains wavefronts with $r_0$ between 8 - 20 cm (in the H band), wind speeds ranging from -15 to 15 m/s in any direction with 3-8 layers, open loop, closed loop and intermediate residuals, photon noise corresponding to between $10^3$ and $10^6$ photons per frame and detector noise between $0$ and $3 \, e^- /pix/frame$. In a single batch, all wavefronts contained different $r_0$, wind speeds and directions, open- and closed-loop residuals, number of photons and detector noise. These values were set following the statistics of the historical observing conditions with the Papyrus instrument \citep{papyrus}.

\begin{figure}
    \centering
    \includegraphics[width=0.99\linewidth]{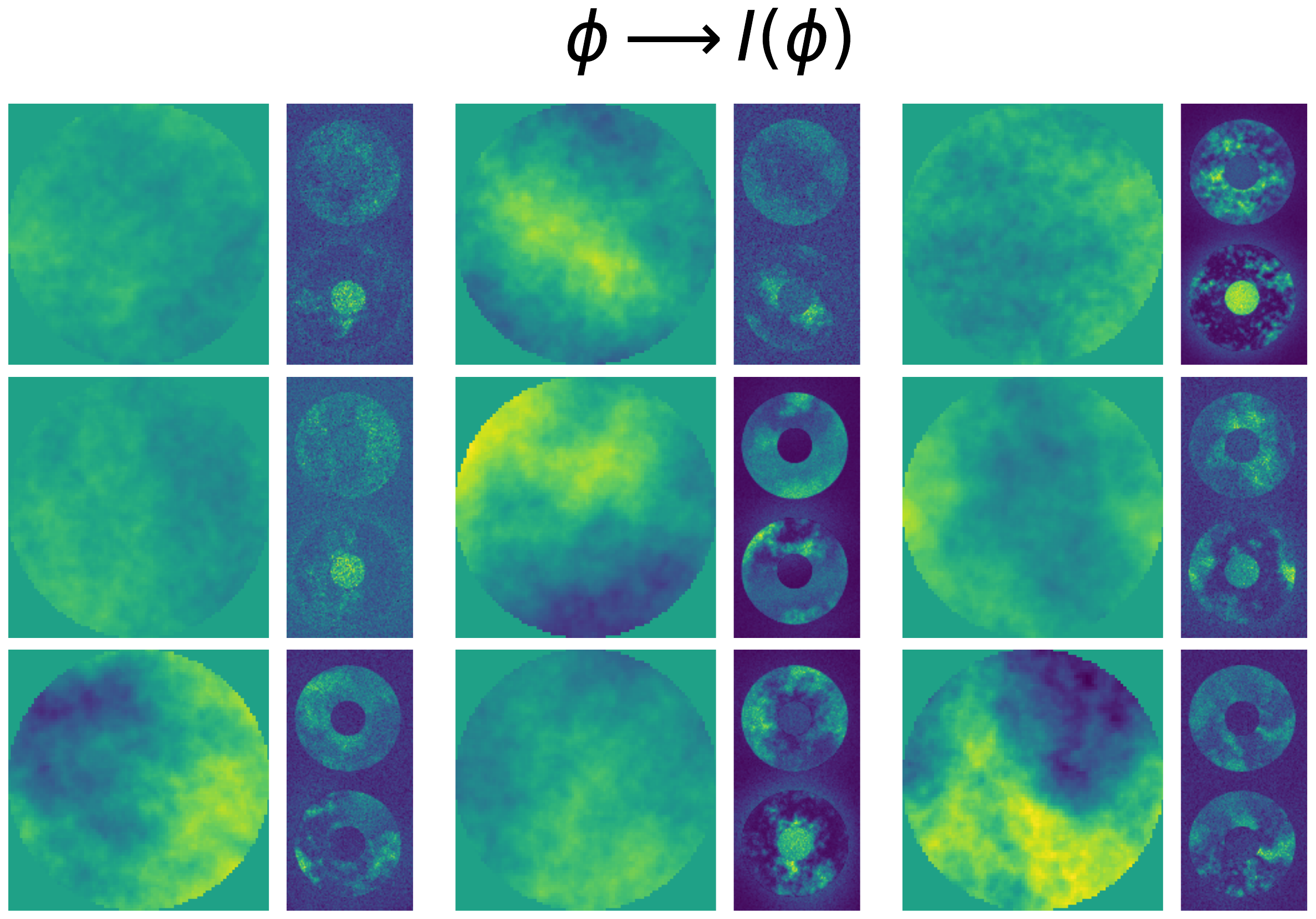}
    \caption{Examples of the training dataset. Each sample consists of a randomly generated atmospheric phase screen $\phi$, and the corresponding v-ZWFS intensity image $I(\phi)$, obtained through the optical model. The same color-scale was used for the phase screens, to highlight their relative strength. Samples span the full range of open-loop, closed-loop and intermediate residual statistics, and used in training.}
    \label{fig:dataset}
\end{figure}

During training, we noticed that the network was prioritizing the accuracy in low order modes and neglecting higher orders. This was because of the energy distribution of the atmospheric PSD, which follows an $f^{-11/3}$ power law. To mitigate this issue we used curriculum training \citep{bengio_curriculum_2009} in which we shifted the training to power laws that better balanced the energy distribution across the modes. We found that by gradually going from $f^{-11/3}$ to $f^{-11/6}$ made the network learn to reconstruct all the modes. Once the network learns to estimate high order modes, going back to $f^{-11/3}$ further improves the overall estimation and it retains its ability to estimate higher-order modes.

To prevent the CNN from learning the exact position of the pupils, which might be slightly different with the instrument, we added a random variation of the pupil positions to each sample. This encouraged the network to not depend on the exact position of these, which proved to be important when deploying the system to the instrument: without it the system would easily diverge from the edges. We found that adding up to 2 pixels of "pupil wandering" drawn from a uniform distribution [-2, 2] during the training gave the best results in terms of stability (i.e. loop convergence on the bench), without any loss in performance.

\subsection{Loss function}

Several loss functions were tested, from the commonly used mean squared error (MSE) and mean absolute error (MAE) like in \citet{2024A&A...687A.202W}, or relative root mean square error (R-RMS) like in \citet{2025A&A...696L...1L}. We found that these worked well to estimate open-loop phases, but the network was unable to perform successfully in closed-loop operation. This occurs because the network, when trained on the full range of open- and closed-loop residuals, is implicitly biased toward larger phase errors which dominate the loss. In AO this is particularly problematic, as the system must perform accurately across a wide dynamic range, from the large aberrations during loop closing to the small residuals at convergence. In other words, the last 1\% of correction can be as important as the previous 99\%, given the exponential decrease in performance with respect to residual phase variance \citep{1983JOSA...73..860M}. To mitigate all of these issues, we propose the use of the mean of the logarithm of the variance of the residual phase $\phi_{res}$ over the batch as the loss function

\begin{equation}
    \mathcal{L} = \frac{1}{n}\sum_{i = 1}^n\log\left[var\left(\phi_{res}\right) \right],
\end{equation}

with

\begin{equation}
    \phi_{res} = \phi_i - \widehat\phi_i,
\end{equation}

\begin{equation}
    \widehat\phi_i = \Phi \cdot \mathrm{CNN}\left(I(\phi_i)\right),
\end{equation}

\begin{figure}
    \centering
    \includegraphics[width=0.99\linewidth]{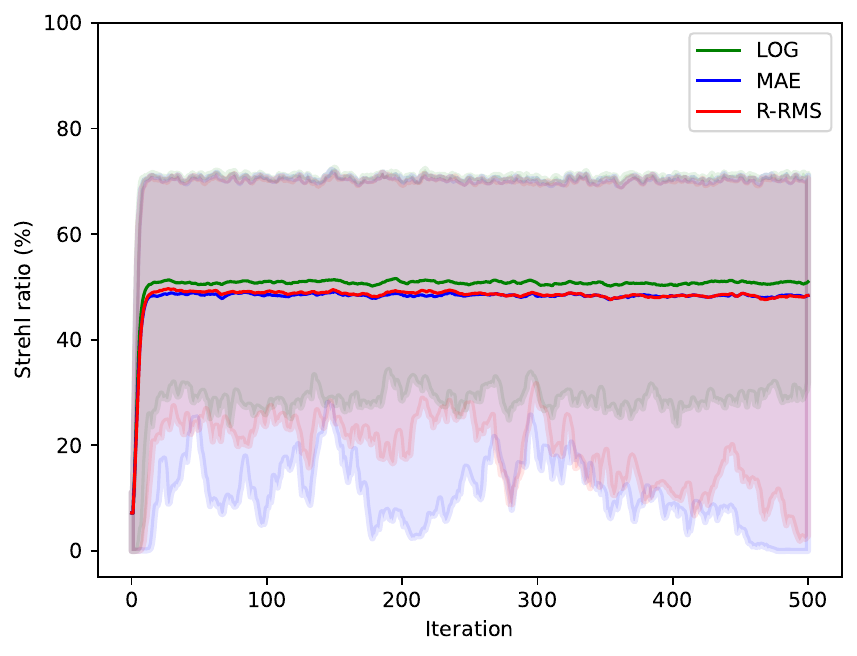}
    \caption{Average SR in the J band for 256 simulated closed-loop runs for each tested loss function. The solid line corresponds to the average SR across all the parallel loops, and the shaded regions correspond to the 25 and 75 percentiles.}
    \label{fig:LossFunctionComp}
\end{figure}

$n$ the batch size, $\Phi$ the matrix that maps the modal coefficients to the DM shape and $CNN(I(\phi_i))$ the output of the CNN given an input frame $I(\phi_i)$. This choice has two important consequences: first, it compresses the high-residual-phase regime, such that a bad inference does not break the training, and second, and most importantly, it expands the low-residual-phase regime, essentially giving the system the physical knowledge of the exponential decrease in performance. To compare the three loss functions under identical conditions, we trained a single CNN using the MAE loss function and then fine-tuned separated copies of it with the MAE, R-RMS, and the proposed log-based loss, ensuring all three variants started from the same initial weights. Note that each loss function is also capable of training a network from scratch; the shared starting point was adopted here solely to control for initialization effects in the comparison. We tested the same 256 atmospheric phase screens in closed loop with each and the results are shown in figure \ref{fig:LossFunctionComp}. The Log-based loss function had the best performance, with an average higher SR and better stability: out of the 256 loops the Log-trained CNN diverged in 22, while both the MAE and R-RMS in 55 and 44, respectively.

\subsection{Training}

We trained the network using the AdamW optimizer \citep{2017arXiv171105101L} with a two-step strategy. In the first step, open-loop training, we generated phase screens on-the-fly, propagated them through the WFS model and computed the loss to perform backpropagation. This step comprised 8000 iterations with batches of 32 images and a learning rate of $10^{-3}$, and was sufficient for the network to learn the general phase reconstruction across the full dynamic range. In the second step, closed-loop training, we simulated the complete AO loop for several iterations per training step (starting from 5 iterations up to 30), with the network correcting its own residual phase at each iteration, exposing it to the dynamic residual statistics it would encounter on-sky. This fine-tuning step comprised 2000 iterations with a reduced learning rate of $10^{-5}$, and was found to be important for stability at the pupil edges, a problem previously identified in \citet{2024A&A...687A.202W}. We found that 30 closed-loop iterations were enough for the network to properly learn real closed-loop residuals and converge in the edges of the pupil. Varying the number of iterations across this range (5 to 30), rather than using a fixed value, exposed the network to residual statistics at different stages of convergence in a controlled manner, from early transient correction to near-equilibrium residuals. In both steps, phase screens, WFS propagation, and noise were generated entirely on-the-fly, with no phase screen, noise realization, or resulting frame ever repeated during training. Both steps also draw from the same underlying phase-screen statistics, comprising open-loop, closed-loop, and intermediate residuals. The full training was performed on an NVIDIA RTX A2000 8GB Laptop GPU and took approximately 30 minutes.

\subsection{Real time computing and data analysis}

The implementation of the AO control was done using the Durham Adaptive Optics (DAO) framework \citep{barr_2025_17264152}. The loop parameters were the same for the linear and CNN reconstructors and they where optimized for each observation. The gain was set to be 0.3 and the leak to be 0.99, with small variations depending on the observing conditions. The linear reconstructor made use of only one of the v-ZWFS pupils, due to instability issues, while the CNN used both of them.

On-sky performance was evaluated from sequences of 5 000 PSF frames acquired at 400 Hz in the J band. Groups of 200 consecutive raw frames were averaged to produce individual PSF estimates, yielding 25 measurements per sequence. The Strehl ratio (SR) of each averaged PSF was computed using the MAOPPY package \citep{2019A&A...628A..99F}, which simultaneously provided an estimate of the seeing. The reported SR for each observation is the median of the 25 individual measurements, providing robustness against short-term fluctuations.

\section{Results}
\label{sec:results}

\subsection{Simulated performance validation}
\label{subsec:simulated_perf}

Before presenting the on-sky results, we first quantify the performance gain of the CNN reconstructor over the linear approach in a controlled simulation environment. Using the same wavefront statistics as those used for training, we generated 5 000 independent phase screens and evaluated both reconstructors in two regimes: open-loop reconstruction and full closed-loop operation. Performance is quantified in terms of residual phase variance, allowing a direct comparison of the two methods across a wide range of input aberration amplitudes. This simulation-based benchmark isolates the reconstruction capability of each method from the operational variability inherent to on-sky observations, and serves as a controlled reference to interpret and support the on-sky results presented in section \ref{subsec:onsky}.

\begin{figure*}
    \centering
    \includegraphics[width=0.45\linewidth]{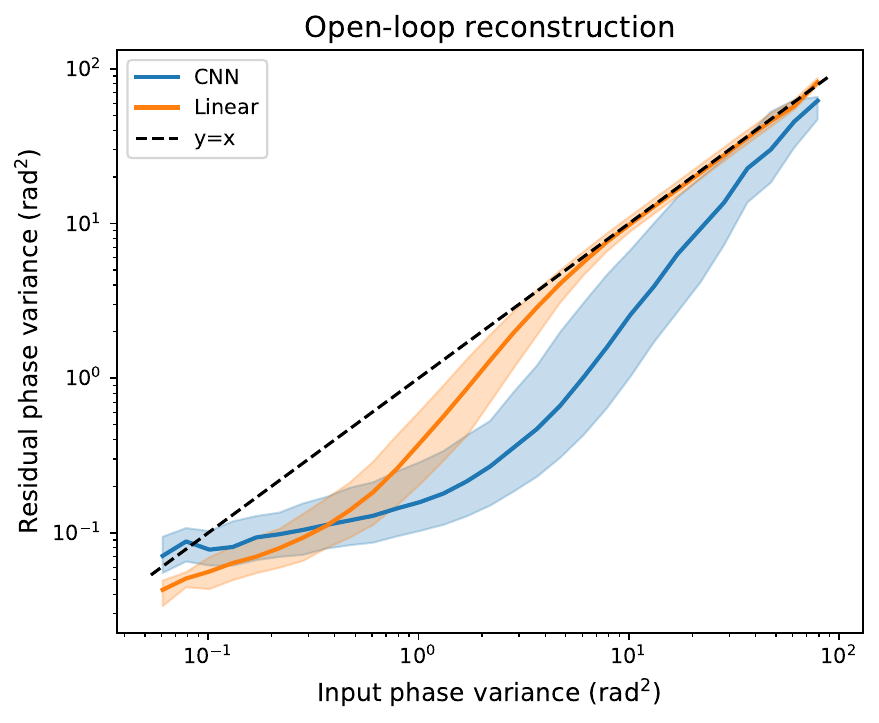}
    \includegraphics[width=0.45\linewidth]{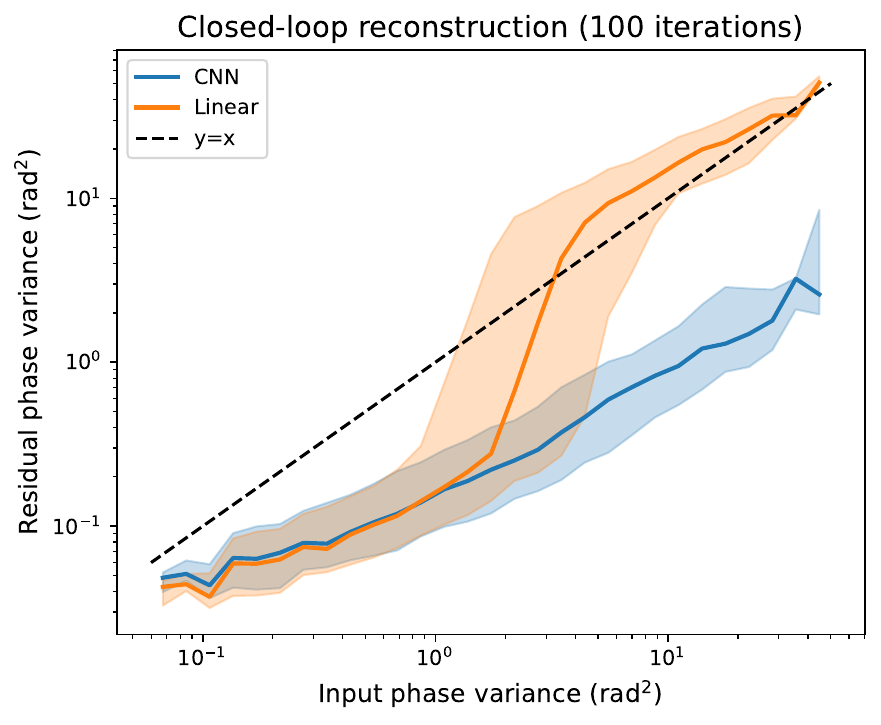}
    \caption{Residual phase variance versus input phase variance for the linear and CNN reconstructors in simulation. Curves represent the median and the shaded regions denote the 10th-90th percentile range. Left panel: residual phase variance after a single reconstruction step; right panel: residual phase variance after 100 closed-loop iterations.}
    \label{fig:InputOutputVariance}
\end{figure*}

Figure \ref{fig:InputOutputVariance} compares the residual phase variance obtained with the linear and CNN reconstructors, for open-loop (left panel) and closed-loop (right panel) operation. The input phase variance follows the same distribution as the training dataset (i.e. open-loop, closed-loop and intermediate residuals). In open-loop reconstruction, the residual variance of the linear method follows the unity line above 5 $rad^2$, indicating that it no longer provides any useful correction, whereas the CNN continues to reduce the residual variance up to input variances of 80 $rad^2$, without yet reaching the unity line, suggesting that closed-loop correction might still be possible. In the intermediate range (0.4-30 $rad^2$), the CNN keeps the residual variance below 50\% of the input, outperforming the linear method. Below 0.4 $rad^2$, the linear reconstructor performs better than the CNN, which is expected since the linear estimator is optimized for this regime, while the CNN is trained across the full dynamic range. Both curves converge toward the unity line below 0.1 $rad^2$, where the residual for either method is dominated by the DM fitting error rather than by reconstruction accuracy. For applications requiring the highest accuracy only at low input phase variance, the CNN could be fine-tuned around that operating point, though this is outside the scope of this work.

In closed-loop operation (right panel in Fig. \ref{fig:InputOutputVariance}), the CNN and linear reconstructors perform similarly for input phase variances between 0.1 and 1 $rad^2$. Above 1 $rad^2$, the residual phase variance of the linear reconstructor progressively approaches the unity line, and no closed-loop run converges for input variances above 5 $rad^2$. In contrast, the CNN keeps the residual variance below 10\% of the input for all tested variances greater than 1 $rad^2$. The successful closed-loop correction for input variances above 30 $rad^2$, where the CNN only partially corrected the phase in open loop, suggests that the network primarily underestimates the phase amplitude rather than its shape or sign. Such gain errors are naturally compensated in closed-loop. Below 0.1 $rad^2$ the linear method slightly outperforms the CNN, consistent with the open-loop trend, although the difference is smaller. In both regimes, the residual floor at low input variance reflects the DM fitting error from correcting only 50 modes.

The simulation results presented above demonstrate that the CNN reconstructor extends the usable capture range well beyond that of the linear method, both in open- and closed-loop operation, while matching or approaching linear performance in the low-residual regime. We now turn to on-sky observations with OZIRIIS to verify whether this extended capture range, established under controlled simulation conditions, translates into a practical operational advantage under real atmospheric turbulence, instrument noise and model imperfections.

\subsection{On-sky demonstration}
\label{subsec:onsky}

\begin{figure*}[t!]
    \centering
    \includegraphics[width=0.99\linewidth]{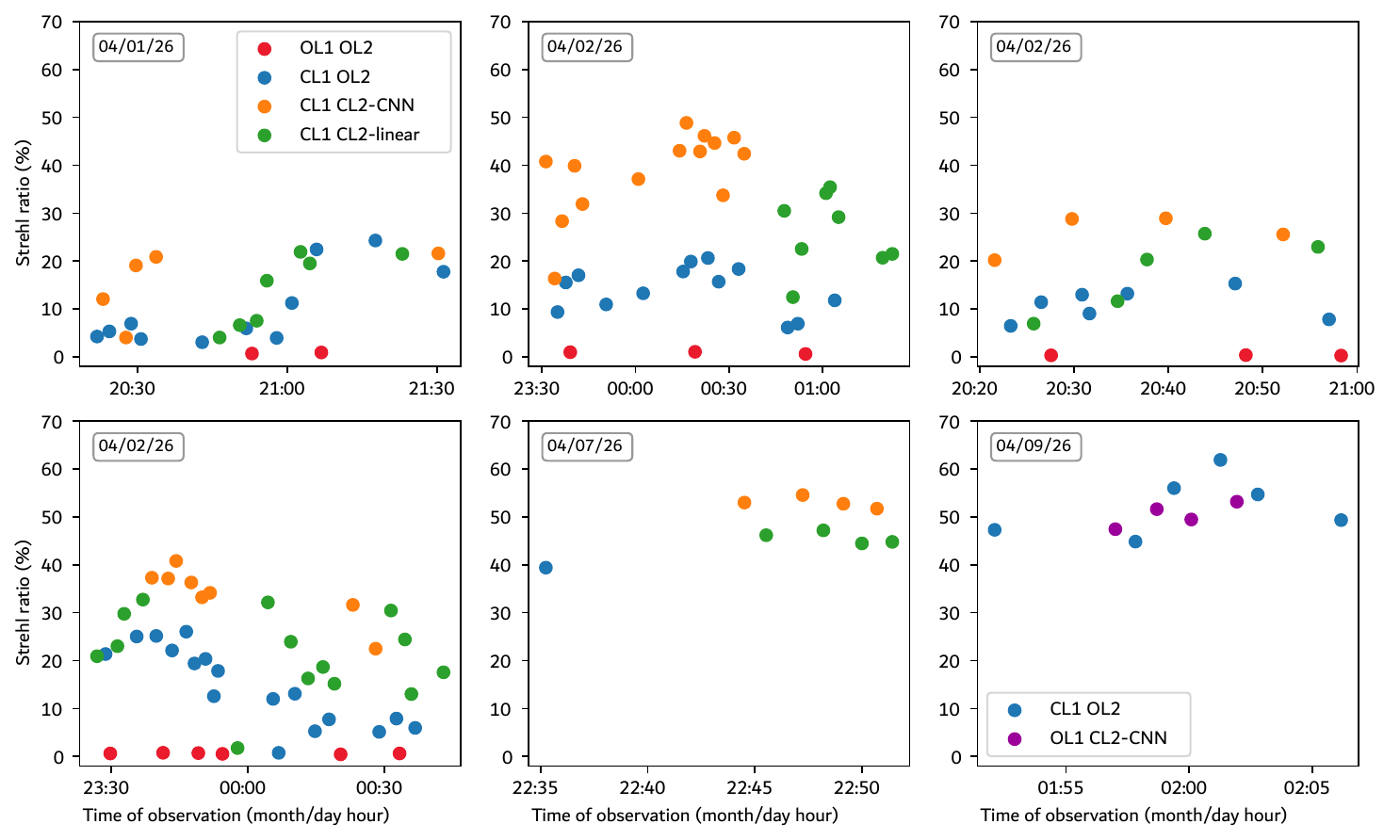}
    \caption{Strehl ratio in the J band obtained during the on-sky observation runs in April 2026. All the plots include the date on the top left corner and the time of the observation is given in UTC time. The plots include five scenarios: open-loop (red); closing the loop only with the first stage (blue); closing the loop with the first and second stage using CNN (orange) and linear (green) reconstructors; closing the loop only with the second stage with the CNN reconstructor (purple).}
    \label{fig:FigureStrehl}
\end{figure*}

\begin{figure*}[ht]
    \centering
    \includegraphics[width=0.9\linewidth]{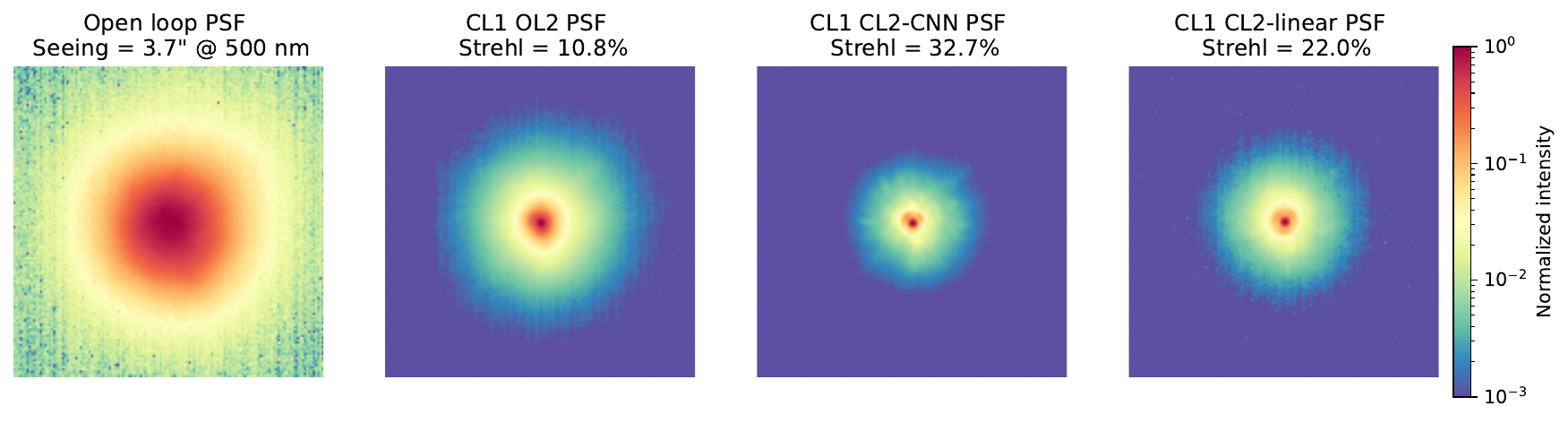}
    \caption{Average PSFs in the J band of the 01/04/2026 - midnight observations (middle-top plot in figure \ref{fig:FigureStrehl}). From left to right: the open loop PSF with a measured seeing of 3.7 arcseconds at 500 nm, the PSF closing only the first stage loop, with a measured SR of 10.8\%, the PSF closing the first and second stage using the CNN reconstructor for the second stage, with a measured SR of 32.7\%, and the PSF closing the first and second stage using the linear reconstructor for the second stage, with a measured SR of 22.0\%}
    \label{fig:FigurePSFs}
\end{figure*}

\begin{figure}
    \centering
    \includegraphics[width=0.99\linewidth]{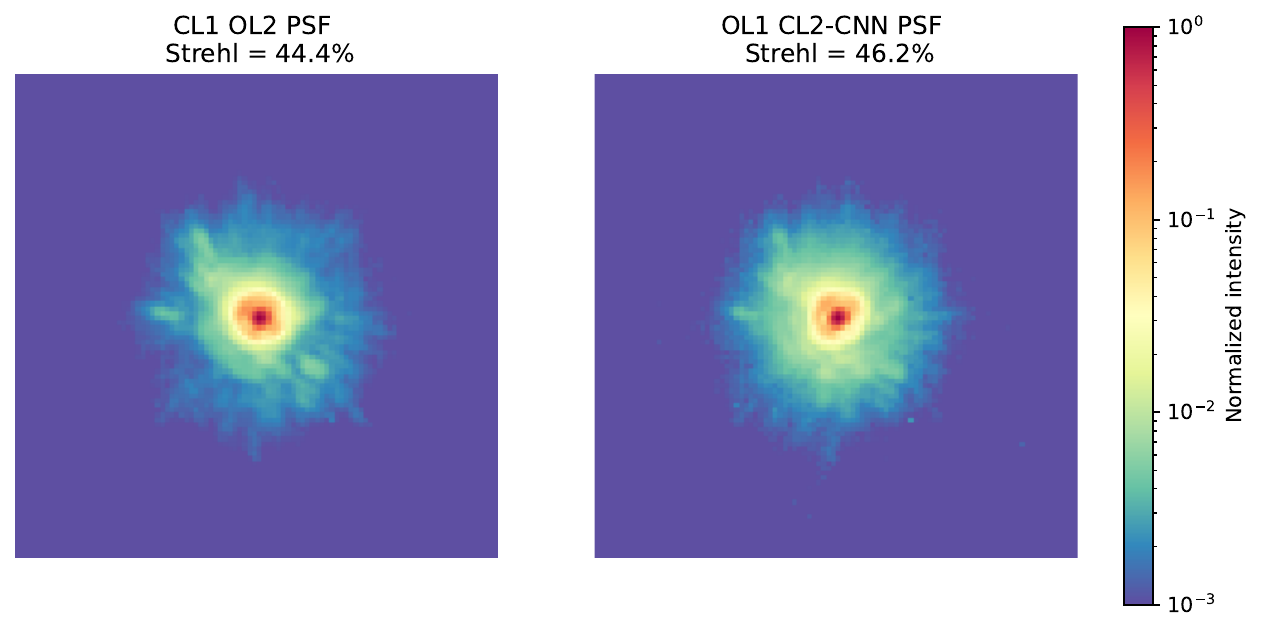}
    \caption{Average PSFs of the 09/04/2026 observations comparing closing only the first or second stage loops. The left PSF was obtained closing only the first stage and a SR of 44.4\% was obtained. The right PSF was obtained closing only the second stage with the CNN reconstructor, in which a SR of 46.2\% was obtained. These observations were taken in quick succession to limit the seeing variability.}
    \label{fig:FigurePSFsComparison}
\end{figure}

Figure \ref{fig:FigureStrehl} shows the Strehl ratio obtained during six observing runs carried out over a week in April 2026. During the observation week, the average seeing was around $3.0 \pm 0.6$ arcseconds at 500 nm. The figure shows five observing modes: open loop (OL1 OL2), closing only the first stage (CL1 OL2), closing both stages with the CNN reconstructor (CL1 CL2-CNN), closing both stages with the linear reconstructor (CL1 CL2-linear), and closing only the second stage with the CNN reconstructor (OL1 CL2-CNN). We note that the dataset reflects real operational usage rather than a controlled benchmark: in several runs only the CNN reconstructor was used, in others both reconstructors were alternated, and in some only the linear reconstructor was tested. Instances where either reconstructor failed to converge were not systematically logged during the campaign and are therefore not individually represented in Fig. \ref{fig:FigureStrehl}. All datapoints shown correspond to runs where the loop successfully converged, with no successful run excluded. The open-loop SR was measured to be less than 1\%: at the observed seeing, the corresponding phase variance is on the order of tens of $rad^2$ or more, far beyond the small-aberration regime where the Maréchal approximation applies \citep{1983JOSA...73..860M}, and well into the regime where the Strehl ratio is expected to be close to zero. Across the campaign, the CNN reconstructor operated successfully over a broader range of conditions than the linear reconstructor, including several runs where the linear approach failed to converge and only the CNN was used. When both reconstructors were operated under comparable conditions, the CNN consistently achieved higher Strehl ratios. A direct comparison of failure rates is not possible, as instances where the reconstructors failed were not systematically recorded; the analysis therefore focuses on performance in conditions where both reconstructors successfully converged. 

Figure \ref{fig:FigurePSFs} illustrates the progressive improvement in PSF quality across the four operating scenarios for the 01/04/2026 midnight observations, under a seeing of 3.7 arcseconds at 500 nm. From left to right: the open loop PSF, the first-stage-only correction (SR = 10.8\%), the first and second stage correction using the linear reconstructor (SR = 22.0\%), and the first and second stage correction using the CNN reconstructor (SR = 32.7\%). The addition of the second stage brings a clear improvement in both cases, but the CNN reconstructor outperforms the linear method, demonstrating the benefit of nonlinear reconstruction under challenging seeing conditions.

The most demanding test of the CNN reconstructor consisted in closing the AO loop using only the second stage, with no upstream correction from the first stage, a regime in which the ZWFS receives the full atmospheric turbulence. During this run, we measured the average seeing to be 2.2 arcseconds at 500 nm, which roughly corresponds to an average input phase variance of 33 $rad^2$ at the operating wavelength (1.5 - 1.7 $\mu m$) of the v-ZWFS \citep{1976JOSA...66..207N}. As shown in the 09/04/2026 panel of Figure \ref{fig:FigureStrehl} and in Figure \ref{fig:FigurePSFsComparison}, this configuration achieved a Strehl ratio of 46.2\%, comparable to the 44.4\% obtained closing only the well-optimized first stage. We note that these observations were taken in quick succession (30 seconds to 1 minute) to minimize seeing variability between tests. This result would be extremely challenging to achieve with the linear reconstructor, as shown in Sec. \ref{subsec:simulated_perf}. We emphasize that although a v-ZWFS was used, which by construction has a higher dynamic range than a traditional ZWFS, it remains subject to the same wrapping artifacts that complicate phase estimation, making this result a genuine demonstration of the extended dynamic range enabled by nonlinear reconstruction, consistent with the expected results from the simulations presented in Fig. \ref{fig:InputOutputVariance}.

Taken together, these results demonstrate that a CNN reconstructor trained entirely in simulation can be successfully deployed on-sky to extend the operational range of the ZWFS well beyond the limits of linear reconstruction. The CNN consistently outperformed the linear reconstructor when both methods converged, and enabled stable closed-loop operation in conditions where the linear approach failed entirely, including, to our knowledge, the first demonstration of closing the loop on full atmospheric turbulence with a ZWFS alone. These findings suggest that nonlinear reconstruction is not merely an incremental improvement over linear methods, but a qualitative shift in the use cases available to highly sensitive but inherently nonlinear wavefront sensors such as the ZWFS.

\section{Conclusion}

In this work, we demonstrated for the first time the on-sky closure of an AO loop using a v-ZWFS with a CNN as phase reconstructor. The CNN, trained entirely in simulation, extended the dynamic range of the v-ZWFS well beyond the limits of linear reconstruction, enabling stable closed-loop operation under conditions where the linear reconstructor failed, and even allowing the v-ZWFS to operate as a first-stage sensor under full atmospheric turbulence. These results were made possible by three key contributions: an accurate instrument model accounting for DM misregistration, a log-variance loss function that encodes the physical importance of small residuals, and a two-step training strategy that ensures closed-loop stability. Together, these demonstrate that a simulation-trained nonlinear reconstructor can be deployed directly on-sky without retraining on real data, a critical requirement for future instruments. We consider these factors, and in particular the fidelity of the instrument model, to be more critical to the on-sky success than the specific CNN architecture, which was adopted once it showed satisfactory performance rather than resulting from an extensive architecture search, leaving further gains through architectural optimization as a promising direction for future work.

More broadly, this work illustrates that the performance limits of AO systems can be significantly extended through the use of nonlinear reconstruction methods. The ZWFS is among the most sensitive wavefront sensors available, but its nonlinearity has historically constrained its use to small-residual phase regimes. By removing this constraint, nonlinear reconstructors make the ZWFS a viable option for a much wider range of observing conditions and instrument configurations. With more sophisticated architectures and faster hardware, further gains are within reach, making this approach particularly relevant for the demanding requirements of the next generation of extremely large telescopes.

%--------------------------------------------------------------------

\begin{acknowledgements}
This work benefited from the support the French National Research Agency (ANR) with the Programme Investissement Avenir F-CELT (ANR-21-ESRE-0008), PEPR ORIGINS (XAO-WFS ANR-22-EXOR-00062 and COMPACT SPECTROGRAPHS ANR-22-EXOR-0006), the ANR-DGA-AID ASTRID program (ANR-25-ASTR-0015), the Action Spécifique Haute Résolution Angulaire (ASHRA) of CNRS/INSU co-funded by CNES, the french government under the France 2030 investment plan (cassiopée project) and the Initiative d’Excellence d’Aix-Marseille Université A*MIDEX, program number AMX-22-RE-AB-151. Authors also would like to thank Kent Wallace and Tobias Wegner for their support on the vector-ZWFS.
\end{acknowledgements}

\bibliographystyle{aa}
\bibliography{bibliography}

\end{document}